# Moving Charge Distributions in Classical Electromagnetism and the FitzGerald-Lorentz Contraction

**Patrick Moylan**

Physics Department, The Pennsylvania State University, Abington College, Abington, PA 19001, USA

E-mail: pjm11@psu.edu

**Abstract.** In [Eur. J. Phys. **25** (2004) 123-126], Dragan V. Redžić is led to the FitzGerald-Lorentz contraction by comparing electromagnetic images of a moving point charge and a moving conducting sphere. We wish to point out that much simpler possibilities intrinsic to electromagnetism already exist from which we may get at the FitzGerald-Lorentz contraction hypothesis. In particular we consider an example going back to Poincaré in [*Bulletin des Sciences mathématiques*, 28, (1904) pp. 302-324] (ref. [2]), in which he considers the problem of two moving, parallel line charges in order to get at length contraction. We develop this model of Poincaré and show that it leads not only to the FitzGerald-Lorentz contraction but also to an elementary derivation of the composition of velocities formula in special relativity for collinear velocities.

Redžić suggests that, by considering such purely electromagnetic examples, the Maxwellians could have been led to the contraction hypothesis much before the time of the Michelson-Morely experiment, and we agree with him that such elementary results as the ones discussed here could not have escaped their attention. Apparently, it took an extremely sensitive experiment, not intrinsic to electromagnetism, such as was the Michelson-Morley experiment, together with the efforts of persons with authority, like Lorentz and Poincaré, trying to uphold the relativity principle before the radical notion of length contraction could seriously be entertained, and making ripe the way for the genius of Einstein.

# 1. Introduction

In a recent article Dragan V. Redžić [1] showed that a study of the properties of certain electromagnetic systems leads directly to the Fitzgerald-Lorentz contraction, and hence opens the door to special relativity. A brief description of his results is as follows. We define an electromagnetic image associated with a given surface charge distribution as a charge distribution contained within one side of a surface and which produces the same electromagnetic field on the other side of the surface as that produced by the surface charge distribution. A familiar example is the following: the electromagnetic image of a spherical conducting shell at rest is a point charge located at its center. Based on calculations contained in his previous papers [3], [4], Redžić states the following two results: (1) the electromagnetic image of a conducting sphere of radius $R$ moving with velocity $\vec{v}$ is a moving line charge of length $2Rv/c$, moving with the same velocity; (2) a point charge moving with velocity $\vec{v}$ is the electromagnetic image of a moving "Heaviside ellipsoid" with semi-axes $R\sqrt{(1-v^2/c^2)}$, $R$, and $R$, where the smallest axis is in the direction of motion. (Redžić implicitly assumes that $v < c$, although the Maxwellians‡ had no reason to believe this.) These two results combined with the principle of relativity applied to two different inertial systems lead to an immediate contradiction, unless the shapes of conducting bodies are affected by their motion. For suppose that a moving, conducting sphere retains its spherical shape when moving, then, according to (1), the image of the conducting sphere is a moving line charge. However, the image of a uniformly moving sphere should be a uniformly moving point charge at its center, since, in passage to the inertial frame in which the sphere is at rest, its image also comes to rest, and must be the point charge, if the principle of relativity is to be upheld. But, according to (2), a uniformly moving point charge is the image of a Heaviside ellipsoid, not of a uniformly moving sphere. Thus we are faced with an incontrovertible paradox, and Redžić concludes that there is no other way out of this situation except to admit that the shape of an object is changed due to its motion, assuming a quantity of charge is not affected by its motion. From this he suggests that the Maxwellians could have arrived at the Fitzgerald-Lorentz contraction hypothesis rather early on by means other than the negative result of the Michelson-Morley experiment (1887) [6], [7], [8], [9].

In this note we wish to point out that much simpler possibilities intrinsic to electromagnetism already exist which lead to the FitzGerald-Lorentz contraction hypothesis. In particular, we consider the example which was used by Poincaré in his 1904 paper [2] outlining much of the foundational aspects of special relativity, namely, two parallel currents i.e. two moving line charges. Following ref. [2] we show that, without the hypothesis of length contraction, it is impossible for the principle of relativity of motion to be upheld in this example of two moving line charges, assuming, as Redžić does, that the quantity of charge on an object is not affected by its motion.

‡ Redžić refers to the main actors in the development of Maxwell's electromagnetic theory as the "Maxwellians" c.f. the book, *The Maxwellians* by B.J. Hunt [5].

## 2. Controversies in Classical Electromagnetism centering around the Relativity of Motion

We start with two parallel currents, more precisely, two infinitely long parallel line charges with identical amounts of charge uniformly distributed on them and separated by a distance $r$ from one another. For concreteness we take the charge on each line charge to be negative, and we assume that the line charges are held at the fixed distance $r$ from one another. In order to achieve this there acts on each line charge a constraining counterforce. This force acts on all parts of each line charge in such a way that the electrostatic repulsion between the two line charges is exactly balanced by this counterforce (or equilibrant force). We give below an explicit description of such a counterforce and describe the force balance produced by this equilibrant force.

From the point of view of an observer moving to the left with speed $v$ relative to the line charges in a direction parallel to the line charges, he sees two parallel currents, i.e. two line charges moving to the right with velocity $\vec{v}$. We analyse the situation in two cases: 1) in the rest frame of the two line charges; 2) in an inertial frame which is moving relative to the line charges, in a direction parallel to them. In case 1, the two line charges repel each other due to their electrostatic repulsion. In case 2, in addition to the same electrostatic repulsion, the moving line charges are equivalent to two parallel currents in the same direction which attract one another. In this case, "the total electromagnetic attraction diminishes and the total repulsion is more feeble than if the two bodies were at rest" [2].

As does Redžić's example, our analysis leading to length contraction hinges upon the veracity of one of the most basic general principles of physics, namely the *principle of relativity of motion*, by which no physical experiment can tell apart a particular inertial frame from any other one. In other words, the principle of relativity of motion is that principle, "according to which the laws of physical phenomena should be the same, whether for an observer fixed, or for an observer carried along in a uniform movement of translation, so that we have not and could not have any means of discerning whether or not we are carried along in such a motion" [2]. We apply this principle together with the assumption that the quantity of charge is not affected by its motion to study the two situations described above i.e. the description of events in two different inertial frames, one at rest relative to the line charges (case 1), and the other moving with velocity $-\vec{v}$ relative to the line charges (case 2). The inertial frame for case 1 is denoted by $S$, and the inertial frame for the second case is $S'$. Fig. 1, depicts the second case i.e. the inertial frame $S'$, relative to which the two line charges are moving to the right with speed $v$. This elementary example is all that we will need in order to get at the conflict between the laws of classical electromagnetism and the relativity principle and to "grasp the implications to physics which upholding the principle of relativity of motion entails" [2], namely, the FitzGerald-Lorentz contraction and its consequences.

In the first case, the inertial frame $S$ in which the line charges are both at rest, there is only the repulsive electrostatic force between the line charges balancing the equilibrant forces on the line charges. Assuming the idealized case of two (essentially) infinitely long line charges, we easily obtain that the magnitude of the net electric field due to one of the line charges at a point on the other line charge is given by

$$E = \frac{\lambda}{2\pi \epsilon_0 r}, \qquad (1)$$

where

$$\lambda = \frac{Q}{L} \qquad (2)$$

is the linear charge density of the line charge, i.e. the ratio of the total charge $Q$ on a length $L$ of a line charge to the length $L$. From this it follows that the net force on a stretch of length $L$ due to the other line charge is

$$F(L) = \int_{-L/2}^{L/2} dq\, E = \frac{\lambda^2}{2\pi\epsilon_0\, r} L = \frac{Q^2}{2\pi\epsilon_0\, r\, L}, \qquad (3)$$

with $dq = \lambda\, dz$ being an infinitesmal amount of charge on an infinitesmal portion of the line charge of length $dz$.

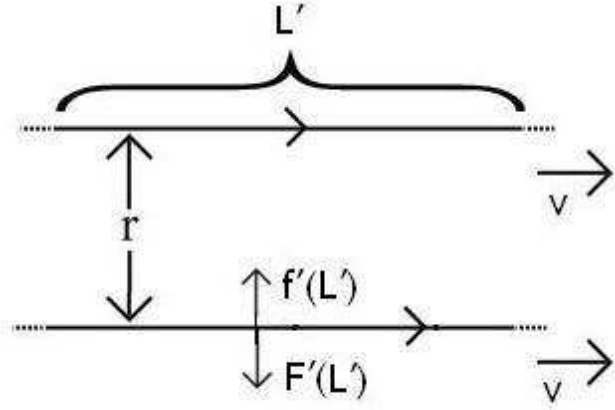

**Figure 1.** Two moving line charges viewed from the reference frame $S'$ in which they are moving to the right with speeds $v$. The electromagnetic repulsion $F'(L')$ of the two moving line charges is balanced by a counterforce $f'(L')$.

In the second case the observer must also detect the magnetic force. Using the *nonrelativistic* Lorentz force law applied to an infinitesimal amount of charge $dq$ of the bottom current, we obtain, by integration, the following for the magnitude of the net electromagnetic force on a stretch of length $L$ of the bottom current due to the top current:

$$F(L) = \left| \int_{-L/2}^{L/2} dq\, (\vec{E} + \vec{v} \times \vec{B}) \right|. \qquad (4)$$

With the help of
$$i = \frac{dq}{dt} = \frac{dq}{dz}\frac{dz}{dt} = \lambda\, v \ , \tag{5}$$
we find
$$B = \frac{\mu_0 \lambda v}{2\pi r} \tag{6}$$
for the magnitude of the magnetic field at the bottom current due to the top current. Substituting this together with eqn. (1) for the magnitude of the electric field at the bottom current due to the top current (say) into eqn. (4), we obtain for the magnitude of the force on a length $L$ of the (moving) bottom line charge:
$$F(L) = \frac{Q^2}{2\pi\epsilon_0\ r\ L} - \frac{\mu_0 Q^2 v^2}{2\pi r\ L}. \tag{7}$$
The reason for the minus sign in front of the second term in right hand side of this equation is, as stated above, because the magnetic force of two parallel currents in the same direction is attractive, and hence the total repulsion is more feeble than in the stationary case. Using
$$\mu_0 \epsilon_0 = \frac{1}{c^2} \tag{8}$$
we simplify this to
$$F(L) = \frac{Q^2}{2\pi\epsilon_0\ r\ L}\left(1 - \frac{v^2}{c^2}\right)\ . \tag{9}$$
Equations (3) and (9) are incompatible with one another. According to the principle of relativity of motion they must be the same, but they are not. We are forced to the conclusion that the length of the moving line charge gets contracted. In our efforts to uphold the principle of relativity, we have been led to the notion length contraction, which first appeared in the bold pioneer work of FitzGerald§ and subsequently by Lorentz, Poincaré, Einstein and others.∥

One could surely devise other mechanisms than the usual one (FitzGerald-Lorentz contraction i.e. eqn. (15)) in order to account for the discrepancy between eqns. (3) and (9). One possibility briefly alluded to by Poincaré in ref. [2] is to consider tampering with the electromagnetic constants $\epsilon_0$ and $\mu_0$, which would be tantamount to supposing that the velocity of light is different for the moving and stationary observer. Another possibility would be to assume that the length $L$ dilates according to some arbitrary power of $\left(1 - \frac{v^2}{c^2}\right)^{1/2}$ different from one, and the force changes by a corresponding power of $\gamma = \left(1 - \frac{v^2}{c^2}\right)^{-\frac{1}{2}}$ in such a way that eqns. (9) and (3) are compatible with one another. However, the arguments presented in the next paragraph together with those given in the next section seem to definitely rule out such a possibility. Thus the principle of relativity along with charge invariance¶ leads most simply to the FitzGerald-Lorentz contraction formula, namely $L' = \gamma^{-1} L$.

§ From his own words on the subject in a letter to Lorentz [10], we know that FitzGerald was ridiculed by some of his colleagues in Dublin for putting forth his idea of length contraction. No wonder that it took persons with the authority of Lorentz and Poincaré to espouse the idea before others would take the notion seriously!
∥ To appreciate just how difficult the strange and perplexing notion of length contraction must have been for the contemporaries of FitzGerald and Lorentz to accept, we point out that even today, more than 120 years after the publication of FitzGerald's paper on the subject issues relating to length contraction are still being addressed. In this regard we mention the so-called "submarine paradox"[11].

We argue as follows: on the basis of the principle of relativity, we should have agreed that the electrostatic force on a stretch of length $L$ in the $S$ frame and the total electromagnetic force on an identical stretch of line charge of length $L$ in the $S'$ frame must be equal.[+] But we have seen that the incompatibility of eqns. (3) and (9) make this impossible. Thus we insist that the length $L$ in the $S'$ frame is contracted. So to a length $L$ in $S$ corresponds a shorter length $L'$ in $S'$ (i.e. for the moving line charge) with identical amounts of charge on both. But we still want to compare eqns. (3) and (9), so the $L$ in eqn. (9) must be for a larger length and thus it must carry a larger amount of charge.

The quantitative details of this approach are as follows. In the $S$ frame the electrostatic force on a stretch of length $L$ is, according to equation (3), given by:

$$F(L) \text{ in } S = \frac{\lambda^2}{2\pi\epsilon_0} \frac{1}{r} L . \tag{10}$$

Similarly, according to equation (9), the force on a contracted stretch of the line charge of length $L'$ in the $S'$ frame is:

$$F(L') \text{ in } S' = \frac{\lambda'^2}{2\pi\epsilon_0 \ r} \left(1 - \frac{v^2}{c^2}\right) L' , \tag{11}$$

¶ We might think to insist that the amount of charge per unit length is different for the moving and stationary line charge while keeping the lengths the same. However, this is seen to be untenable on account of *charge quantization*, something which surely was not appreciated by the Maxwellians. To see this, we take the current $i$ to be so tiny and the charge on an individual charge carrier to be so small (for instance, if the individual charges were electrons) that there is just one unit of electron charge in one charge carrier flowing across a given unit of length at any any given instant. But this would mean, for the situation in the stationary frame $S$ that a fractional amount of charge, in general, incommensurate with the electron's charge, would be across the unit of length. Such a situation clearly violates the principle of charge quantization. Note, however, that for the situation just described, eqn. (1) would only have approximate validity. Eqn. (1) applies to constant linear charge density, and not for electrons viewed as spherical charges separated by considerable distances from one another, as would have to be the case for the argument given here to be valid.

[+] Suppose that equations (3) and (9) were different, and imagine that the two line charges are being held in place by an equilibrant force acting in the direction perpendicular to the line charges. Assume that the force which the moving observer observes on a unit of length is barely less than the force necessary to overcome this equilibrant force. Then, assuming that the equilibrant force is the same for both observers, according to the observer who is stationary, the force on a unit of length is greater than this equilibrant force, and there would be a net force on the line charge, so that it would start to move in the perpendicular direction! This situation is clearly untenable, if the principle of relativity of motion is to be upheld. It cannot be that the line charge moves in the perpendicular direction in the $S$ frame and yet be stationary in the $S'$ frame.

where $\lambda'$ equals $\frac{Q}{L'}$, and, likewise, the force on a stretch of length $L$ in the $S'$ frame is

$$F(L) \text{ in } S' = \frac{\lambda'^2}{2\pi \epsilon_0 r} \left(1 - \frac{v^2}{c^2}\right) L \ . \tag{12}$$

Now equating equations (10) and (12) we get

$$F(L) \text{ in } S = F(L) \text{ in } S' \tag{13}$$

implies

$$\lambda'^2 \left[1 - \frac{v^2}{c^2}\right] = \lambda^2 \ , \tag{14}$$

and, from the definitions of $\lambda$ and $\lambda'$, we obtain

$$L' = \left[1 - \frac{v^2}{c^2}\right]^{1/2} L, \tag{15}$$

which is the FitzGerald-Lorentz contraction formula.

Notice that our just obtained result implies

$$\frac{F(L)}{L} \text{ in } S = \frac{\lambda^2}{2\pi \epsilon_0 r} = \frac{\lambda'^2}{2\pi \epsilon_0 r}\left(1 - \frac{v^2}{c^2}\right) = \frac{F(L')}{L'} \text{ in } S' \ . \tag{16}$$

Instead of designating the force on a given length $L$ in $S$, and the corresponding force on the corresponding contracted length $L'$ in $S'$ by $F(L)$ in $S$ and $F(L')$ in $S'$, respectively, let us simply denote them by $F(L)$ and $F'(L')$. Then the just given equation can be rewritten as

$$\frac{F'(L')}{L'} = \frac{F(L)}{L} \ . \tag{17}$$

This leads immediatley to the correct transformation law of force in the direction transverse to the direction of motion, namely: [12]

$$F'(L') = F(L) \frac{L'}{L} = F(L) \left(1 - \frac{v^2}{c^2}\right)^{1/2} \ . \tag{18}$$

Specifically, this means that if $F(L)$ is the force on a unit of length $L$ in the rest frame $S$ on the line charge, then the force $F'(L')$ is the force on the same unit of length measured by an observer who sees $L$ moving and hence contracted in length to $L'$. So by the hypothesis of length contraction we have satisfied the requirement that the force on a unit of length $L$ in the rest frame of the line charges and the force on the identical unit of length $L$ in the moving frame (which corresponds to a larger unit of length in the rest frame) are the same, and, yet at the same time we have satisfied the requirement of electromagnetism that the total electromagnetic force on a given unit of length is diminished in the moving frame.

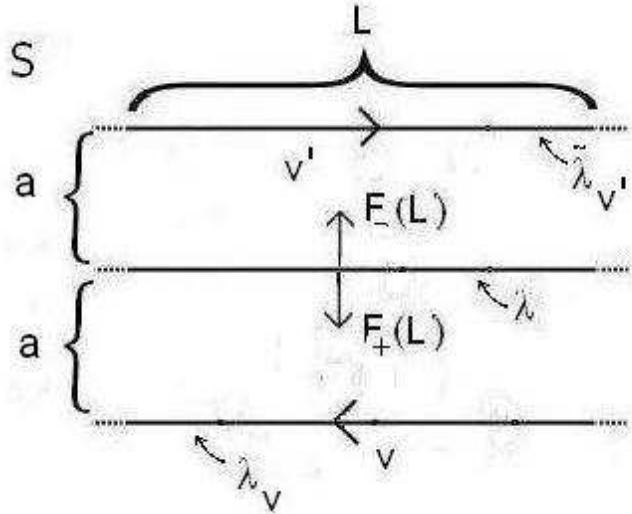

**Figure 2.** Three parallel line charges viewed from the rest frame $S$ of the middle line charge. The top line charge is moving to the right with speed $v$ and the bottom line charge is moving to the left with speed $v$. The electromagnetic repulsion $F_+(L)$ between the top and middle line charges exactly balances the electromagnetic repulsion $F_-(L)$ between the bottom and middle line charges.

## 3. The FitzGerald-Lorentz Contraction Factor and Addition of Velocities

By upholding the principle of relativity we are forced to the necessity of the equality of eqns. (3) and (9), which we have recast in the form of eqn. (13) i.e.

$$F'(L) = F(L) . \qquad (19)$$

Using this we were led, via our particular interpretation, in a straightfowrard manner, to eqn. (14). However, as stated above, there are more more interpretations compatible with eqn. (13) (i.e. eqn. (19)). In general, we may assume that the length dilates according to some arbitrary function of $v$ and the force changes by some corresponding function of $v$ in such a way that for some length $\tilde{L}$ we have that $F'(\tilde{L}) = F(L)$ holds.

We now present arguments to dispense with these other possibilities. Let $L$ be the length of a section of a line charge in the frame $S$ and let $L'$ be its length in $S'$. Then $L$ and $L'$ are related by

$$L' = k(v) L \qquad (20)$$

where the FitzGerald-Lorentz contraction factor $k(v)$ is now assumed to be any arbitrary function of $v$. Consider the three line charges as shown in the Fig. 2. The separation distance between the middle line charge and either one of the other two is $a$. We designate by $S$ the rest frame of the middle line charge. In $S$ the upper line charge and the lower line charge move with speed $v'$ to the right and speed $v$ to the left, respectively.

Let $Q_0$ be the charge on a length $L$ of the middle line charge, then

$$\lambda = \frac{Q_0}{L}. \tag{21}$$

We assume that charge on a length $L$ of the lower line charge as measured in the rest frame of the lower line charge is also $Q_0$, and we choose the charge $\tilde{Q}_0$ on a length $L$ of the upper line charge at rest to be such that the force $F_+(L)$ on a length $L$ of the middle line charge due to the upper line charge exactly balances the force $F_-(L)$ on a length $L$ of the middle line charge due to the lower line charge in the frame $S$. Assuming that the quantity of charge is not affected by its motion, the line charge densities of the lower line charge and upper line charge must, respectively, be

$$\lambda_v = \frac{Q_0}{L_v} \tag{22}$$

and

$$\tilde{\lambda}_{v'} = \frac{\tilde{Q}_0}{L_{v'}} \tag{23}$$

with

$$L_v = k(v)L \tag{24}$$

and

$$L_{v'} = k(v')L . \tag{25}$$

Using eqns. (1) and (2) force balance in $S$, i.e. $F_-(L) = F_+(L)$, leads to

$$\frac{Q_0}{2\pi\epsilon_0} \frac{\lambda_v}{a} = \frac{Q_0}{2\pi\epsilon_0} \frac{\tilde{\lambda}_{v'}}{a} \tag{26}$$

which implies

$$\tilde{\lambda}_{v'} = \lambda_v . \tag{27}$$

Using the definitions of $\tilde{\lambda}_{v'}$ and $\lambda_v$ this leads to

$$\tilde{Q}_0 = \frac{L_{v'}}{L_v} Q_0 . \tag{28}$$

Now consider the situation from the rest frame $S'$ of the lower line charge. In this frame the lower line charge is at rest, the middle line charge moves to the right with speed $v$ and the upper line charge moves to the right with speed $u$ as shown in Fig. 3. Let $F'_-(L')$ ($F'_+(L')$) be the force due to the lower (upper) line on a contracted length $L'$ of the middle line charge. On the basis of the principle of relativity, force balance in $S$ implies force balance in $S'$ so that

$$F'_-(L') = F'_+(L') . \tag{29}$$

Using eqns. (1), (2), (5) and (6) we obtain

$$F'_-(L') = \frac{Q_0}{2\pi\epsilon_0} \frac{\lambda}{a} \tag{30}$$

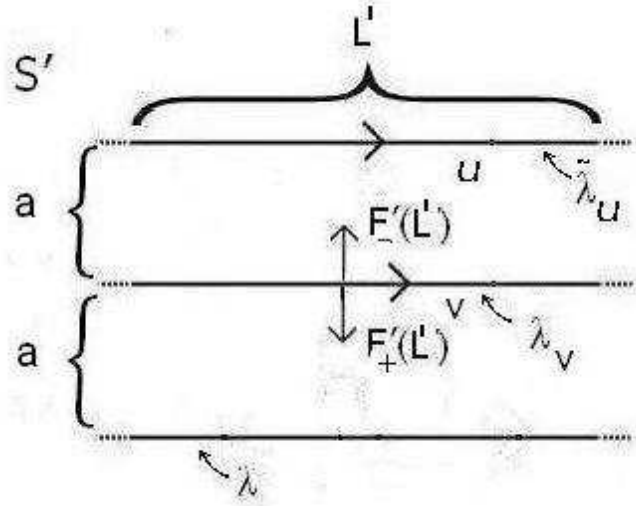

**Figure 3.** Three parallel line charges of Fig. 2 viewed from the rest frame $S'$ of the bottom line charge. In $S'$ the top line charge is moving to the right with speed $u$ and the middle line charge is moving to the right with speed $v$.

and
$$F'_+(L') = \frac{Q_0}{2\pi\epsilon_0}\frac{1}{a}\left(\tilde{\lambda}_u - \frac{vu}{c^2}\tilde{\lambda}_u\right) = \frac{Q_0}{2\pi\epsilon_0}\tilde{\lambda}_u\left(1 - \frac{v\,u}{c^2}\right) \qquad (31)$$
where
$$\tilde{\lambda}_u = \frac{\tilde{Q}_0}{L_u}\,. \qquad (32)$$
Equating eqns. (30) and (31) leads to
$$\lambda = \tilde{\lambda}_u\left(1 - \frac{vu}{c^2}\right), \qquad (33)$$
which gives
$$\frac{Q_0}{L} = \frac{\tilde{Q}_0}{L_u}\left(1 - \frac{vu}{c^2}\right) = \frac{Q_0\,L_{v'}}{L_v L_u}\left(1 - \frac{vu}{c^2}\right) \qquad (34)$$
where in obtaining the last equality we have used eqn. (28). From this equation we obtain
$$L_v L_u = L\,L_{v'}\left(1 - \frac{vu}{c^2}\right) \qquad (35)$$
or
$$k(v)k(u) = k(v')\left(1 - \frac{vu}{c^2}\right)\,. \qquad (36)$$
For $v' = 0$ we must have $u = v$ and, in this case, the just derived equation reduces to
$$k(v)^2 = 1 - \frac{v^2}{c^2}\,. \qquad (37)$$

Hence we must have

$$k(v) = \sqrt{1 - \frac{v^2}{c^2}} \qquad (38)$$

for the FitzGerald-Lorentz contraction factor. This completes our argument to dispense with the other possibilities.

Using our just obtained results it is easy to get at the composition law for collinear velocities in special relativity. We substitute our result, eqn. (38), into eqn. (36) and get the following:

$$\sqrt{\left(1 - \frac{v^2}{c^2}\right)\left(1 - \frac{u^2}{c^2}\right)} = \left(1 - \frac{vu}{c^2}\right)\sqrt{1 - \frac{v'^2}{c^2}} \,. \qquad (39)$$

Solving this equation for $u$ we find

$$u = \frac{v + v'}{1 + \frac{vv'}{c^2}} \qquad (40)$$

which is the relativistic addition of velocities formula for collinear motion [13].

## 4. Justification of Force Balance and Moving Line Charges of Finite Length

It turns out that there is no experimentally measurable difference at all between respective forces in the moving and stationary frames of reference, even though the total electromagnetic force is diminished in the $S'$ frame. Poincaré seems to have been fully aware of all of this as early as 1904, for exactly such is stated in the above mentioned paper of his ([2]): it is true that the magnetic force due to the currents diminishes the total electromagnetic force, but any other force in the same direction is "reduced in the same proportion, (and) we perceive nothing" ([2]).

In order to fully understand Poincaré's assertion, we need to make more precise the counterforces on the two line charges holding them in their positions at a fixed distance $r$ from one another, and to compare transformation laws for the equilibrant forces in the $S$ and $S'$ frames with eqn. (18). For this purpose we imagine that the system of the two line charges in equilibrium with an assembly of fixed springs attached to the line charges at regular intervals as shown in Fig. 4.

In ref. [14] is described a remarkably clever arrangement of springs attached to infinitely long rods which are permitted to move relative to one another via a mechanical construction inolving frictionless ball bearings sliding along the middle rod. It is shown in ref. [14] that this arrangement provides us with a means of giving an empirical and purely mechanical derivation of the force transformation law in special relativity, making use only of the above enunciated principle of relativity and symmetry considerations together with the formula of length contraction, i.e. eqn. (15). We briefly describe the analysis of ref. [12] for the case relevant to us i.e. the case in which the force acts transverse to the direction of motion. A rod $D$ is located exactly in the middle of two other rods, all rods being identical in composition and shape. Rod $D$ is acted upon by spring forces produced by identical springs attached to the other rods as shown in Fig. 5. Each spring is fastened at equal distances to the rods, and the springs are attached to the middle rod by frictionless ball bearings that slide at $D$'s surface. In the rest frame of $D$, the system rod $A$ + springs $A$ and the system rod $B$ + springs $B$ travel in opposite directions with equal speeds $u$ relative to $D$ (c.f. Fig. 5). Thus the net force acting upon a section of length $D$ of the middle rod is zero, and consequently $D$ has zero acceleration in its rest frame.

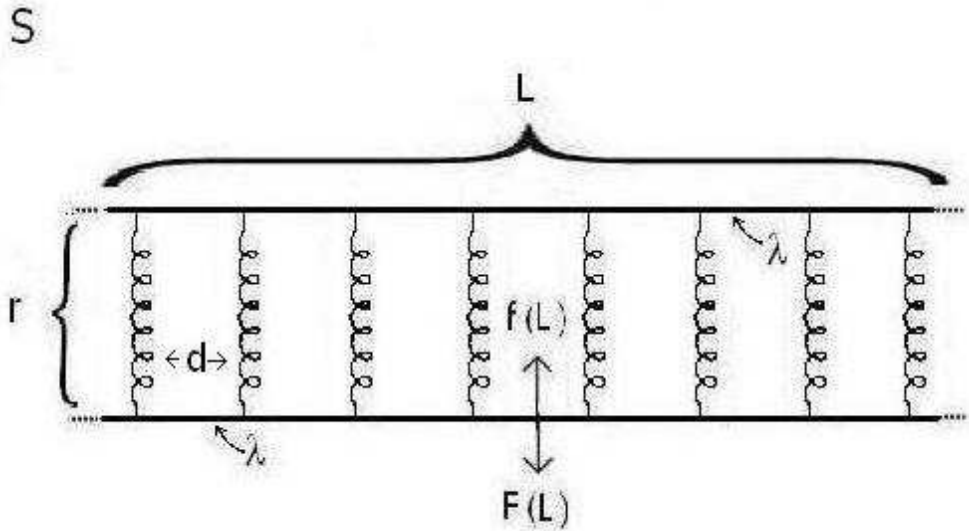

**Figure 4.** The two line charges in static equilibrium with a system of springs as viewed in the reference frame $S$. The ends of the springs are attached to the line charges as shown, and in $S$ the distance between adjacent springs is $d$. The spring force of an individual spring is $f_0$ and the total spring force on a length $L$ is $F(L) = n f_0$ where $n$ is the number of springs in the length $L = n\, d$. (The distance from and endpoint of the line segment $L$ to the nearest spring is $d/2$.)

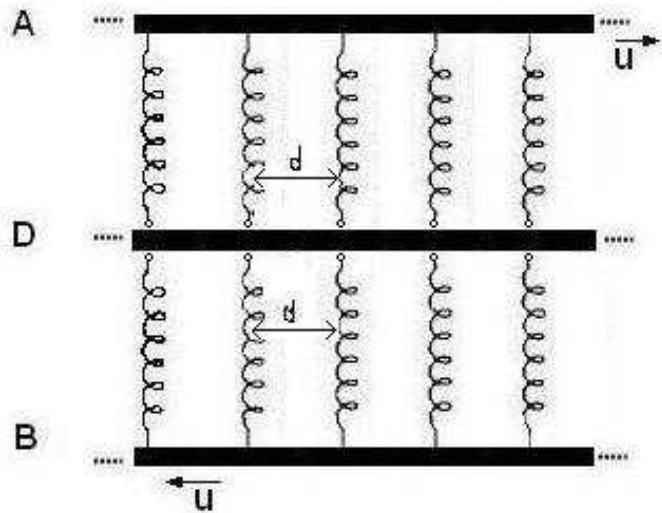

**Figure 5.** A system of three infinitely long rods, $A$, $D$ and $B$, connected by identical springs, with the middle rod $D$ being connected to the springs via frictionless pulleys. The upper and lower rods move relative to the rod $D$ with equal speeds $u$ in opposite directions, and they are held fixed in place, so that the upper and lower springs push or pull equally on the middle rod.

Now consider the situation relative to rest frame of rod B as shown in Fig. 6. Relative to $B$ the system rod $A$ + springs $A$ travels to the right with speed $v$, and the distance between adjacent springs of the system $A$ is

$$d_A = d_B \left(1 - \frac{v^2}{c^2}\right)^{1/2} . \tag{41}$$

Furthermore, relative to $B$ the number of $B$ springs in a section of length $L$ of rod $D$ is

$$n_B = L/d_B \tag{42}$$

and the number of $A$ springs in a section of length $L$ of rod $D$ is

$$n_A = L/d_A . \tag{43}$$

By the principle of relativity the net forces produced by the springs on the section of length $L$ of $D$ must be zero, since it is so in the rest frame of $D$. Hence

$$n_B f_B = n_A f_A \tag{44}$$

where $f_B$ and $f_A$ are the individual spring forces acting upon $D$ by the systems $B$ and $A$, respectively, and using eqns. (41), (42) and (43) we obtain

$$f_A = f_B \left(1 - \frac{v^2}{c^2}\right)^{1/2} . \tag{45}$$

From this we conclude that if $f_B$ is the force exerted by a stationary spring, then $f_A$ as given by eqn. (45) is the force exerted by an identical spring moving moving with speed $v$ in a direction transverse to the length of the spring.

Let us now make use of eqn. (45) to show explicitly that force balance in $S$ implies force balance in $S'$, and hence justifying Poincaré's remarks. From the above and from Figures 4 and 7 we have

$$f'(L') = n f_A = n f_B \left(1 - \frac{v^2}{c^2}\right)^{1/2} = f(L) \left(1 - \frac{v^2}{c^2}\right)^{1/2} . \tag{46}$$

By comparing this equation with eqn. (18) we obtain

$$f'(L') = F'(L') \iff f(L) = F(L) . \tag{47}$$

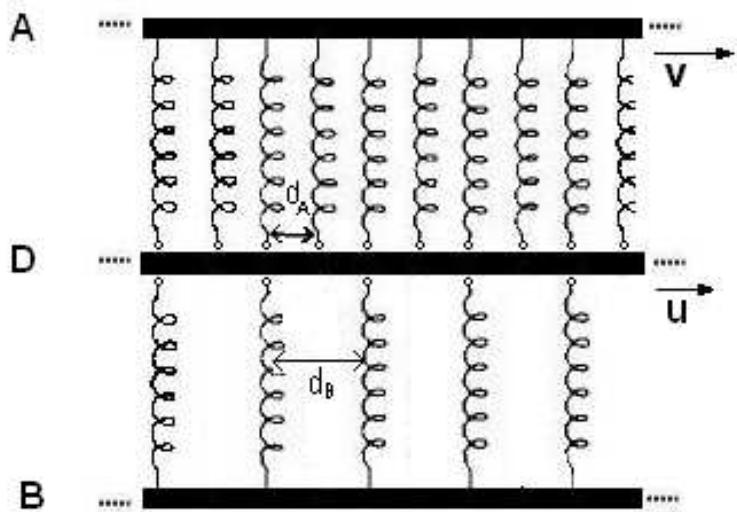

**Figure 6.** The system of rods and springs of Fig. 5 shown in rest frame of the bottom rod, $B$. Relative to $B$ the middle rod moves with speed $u$ and the upper rod moves with speed $v$. The distance between the springs of the system $A$ (upper rod) undergoes length contraction, and consequently the force produced by an individual spring of the system $A$ must decrease in the same proportion as the length.

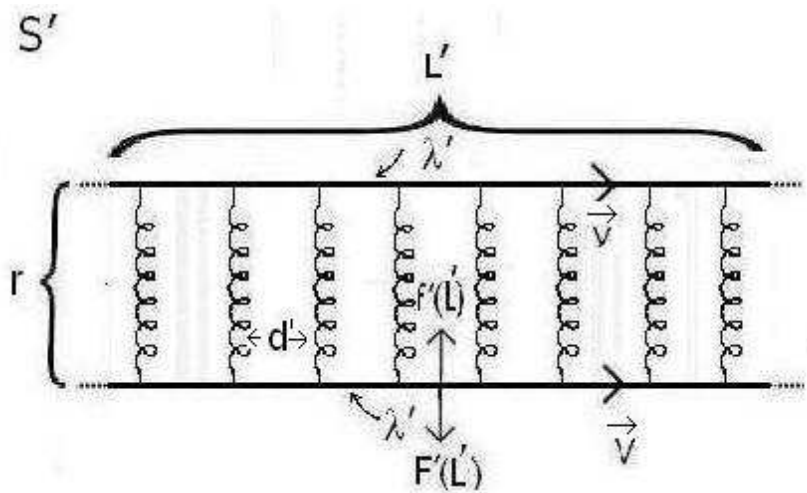

**Figure 7.** The two line charges in kinetic equilibrium with a system of springs as seen in reference frame $S'$. The line charges and the springs all move the right with speeds $v$. The spring force of an individual spring is now $f_0'$. There are $n$ springs in the contracted length, $L'$, so that the distance between adjacent springs is now $d' = d\left(1 - \frac{v^2}{c^2}\right)^{1/2}$.

Hence, if $f(L)$ exactly balances the electrostatic repulsion in the $S$ frame, then $f'(L')$ exactly balances the net electromagnetic force in the the $S'$ frame, and visa versa.

In the above analysis we have considered infinite line charges i.e. we have neglected end effects for line charges of finite length, and it would be useful to carry out a similar analysis for line charges of finite length. There are well-known elementary results for static line charges and stationary currents of finite length i.e. analogs of eqns. (1) and (6) for static line charges and stationary curents of finite length, respectively, e.g. [16]; however, what we need are corresponding results for moving line charges and moving currents of finite lengths. Information pertaining to this problem is used by Redžić himself in his arguments to get at length contraction. For example, he uses in [1] the result that the electromagnetic image of a conducting sphere of radius R, uniformly moving with speed $v$, is a moving line charge of length $2Rv/c$ [4]. Furthermore, in the limit $L \to \infty$, eqns. (3) and (6) hold quite generally in all inertial reference frames, for it may be shown that both Gauss' law and Ampère's law in integral form apply to charges and currents moving at constant velocities relative to their rest frames (c.f. [15]). All of this suggests that results for line charges of finite length should carry over to ours in the limit of very long line charges.

In order to see that this is indeed so and to obtain precise results for end effects we start with the exact expressions for the electric and magnetic fields of a point charge at rest and for a uniformly moving point charge. We then obtain exact expressions for the force between two interacting line charges of finite lengths by suitable integrations, and show that for $L$ large the exact expressions reduced to eqns. (3) and (9). For a point charge at rest, having infinitesmal charge $dq$ there is only an electric field and it is just the Coulomb field, given by

$$d\vec{E} = \frac{1}{4\pi\epsilon_0} \frac{dq\,\vec{r}_0}{r_0^3} \qquad (48)$$

where $\vec{r}_0$ is the displacement vector from the charge to the field point. Using this eqn. we can show that eqn. (3) follows by integration and then by taking the limit of large $L$. We see this as follows. A straightforward consequence of eqn. (48) is the following formula for the electric field of a uniformly charged rod of finite length $L$ and total charge $Q$ with $\lambda = Q/L$, namely: [16]

$$\vec{E}(P) = E_r\,\hat{r} + E_z\,\hat{k} \qquad (49)$$

with

$$E_r = -\frac{\lambda}{4\pi\epsilon_0}\left(\frac{\cot\theta_2}{r_2} + \frac{\cot\theta_1}{r_1}\right) \qquad (50)$$

and

$$E_z = \frac{\lambda}{4\pi\epsilon_0}\left(\frac{1}{r_2} - \frac{1}{r_1}\right) \qquad (51)$$

where the point $P$ does not lie along the axis of the rod and $r_1$ and $r_2$ are the distances from the left end of the rod and the right end of the rod to the point $P$, respectively, and $\theta_1$ and $\theta_2$ are the exterior angles formed at the corresponding two vertices of the triangle formed by the line segment of length $L$ which is the rod, and the point $P$, and which are not at $P$ (see Fig. 8). ($\hat{r}$ and $\hat{k}$ are two of the three standard unit vectors for cylindrical coordinates, the other unit vector, $\hat{\phi}$, being in the azimuthal direction[15].) Using eqn. (30) we obtain for the net force on one line charge due to the other:

$$F(L) = \int_{-L/2}^{L/2} dq \; E_r = -\frac{\lambda^2}{4\pi\epsilon_0} \int_{-L/2}^{L/2} dz \left( \frac{\cot\theta_2}{r_2} + \frac{\cot\theta_1}{r_1} \right) \tag{52}$$

which is an expression valid for parallel line charges of finite length $L$ separated by a distance $r$ from one another. (It is straightforward to see from symmetry that $F_z = \int_{-L/2}^{L/2} \lambda \, dz \, E_z = 0$, and thus the net electromagnetic force between the two rods acts only along the direction of their perpendicular bisector, paralleling the infinite case treated above.) From Fig. 8 (in the rest frame) we have $r_1 = \sqrt{(\frac{L}{2} + z)^2 + r^2}$ and $r_2 = \sqrt{(\frac{L}{2} - z)^2 + r^2}$. Also $\cot\theta_1 = -\frac{\frac{L}{2}+z}{r}$, $\cot\theta_2 = -\frac{\frac{L}{2}-z}{r}$ and substitution of these equations into (52) give

$$F(L) = \frac{\lambda^2}{4\pi\epsilon_0} \int_{-L/2}^{L/2} dz \left( \frac{\frac{L}{2} - z}{r \, r_2} + \frac{\frac{L}{2} + z}{r \, r_1} \right). \tag{53}$$

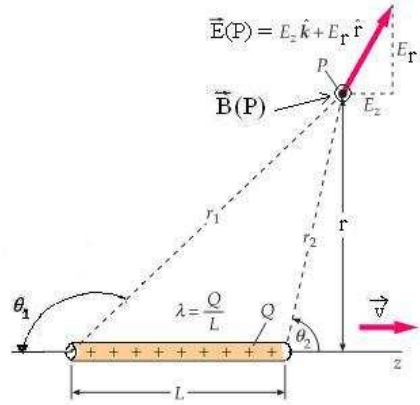

**Figure 8.** A moving line charge of finite length $L$ and charge $Q$ and its electric and magnetic fields. (The line charge is moving to the right with speed $v$.)

We perform the integration to obtain

$$F(L) = \frac{\lambda^2}{4\pi\epsilon_0 \, r} \left( 2\sqrt{L^2 + r^2} - 2r \right) = \frac{\lambda^2}{2\pi\epsilon_0 \, r} L \left( \sqrt{1 + r^2/L^2} - r/L \right) \tag{54}$$

which to lowest order in $r/L$ is exactly eqn. (3).

Now we turn to the more difficult case of two moving parallel line charges of finite lengths $L$. For a uniformly moving point charge, $q$, having velocity $\vec{v}$, we obtain for $\vec{E}$ and $\vec{B}$ the following: [15], [17], [18]

$$\vec{E} = \frac{1}{4\pi\epsilon_0} \frac{q\,\vec{r}_0\left(1 - \frac{v^2}{c^2}\right)}{r_0^3\left(1 - \frac{v^2}{c^2}\sin^2\theta\right)^{3/2}} \quad , \tag{55}$$

$$\vec{B} = \frac{1}{c^2}\,\vec{v}\,\times\,\vec{E}, \tag{56}$$

where $\vec{r}_0$ is the displacement vector from the charge $q$ at the time $t$ to the field point at time $t$, and $\theta$ is the angle between $\vec{v}$ and $\vec{r}_0$. (These equations follow by straightfoward computation from the Liénard-Wiechert potentials for a uniformly moving point charge moving with speed $v$, and already as early as 1867 Riemann and then Lorenz, following Riemann's ideas, had introduced equations similar to the Liénard-Wiechert potentials (ref [18], page 268). Thus, although Liénard-Wiechert's papers, which introduced their famous potentials, were published some years after the Michelson-Morely experiment, i.e. circa 1900, it seems that much about them was already quite well appreciated by the Maxwellians years earlier.)

Consider two line charges each having length $L$, one directly above the other, and separated by a distance $r$ from each other. (The figure depicting this would be Fig. 8, except the top line charge is not shown). Now we follow the same reasoning as after eqn. (3) for moving line charges leading to eqn. (9), except that now the line charges are of finite length. Let $dq' = \lambda dz'$ be an infinitesmal amount of charge on the top line charge and let $d\vec{E}$ and $d\vec{B}$ be the electric and magnetic fields due to an infinitesmal amount of charge $dq = \lambda dz$ of the bottom line charge. $d\vec{E}$ and $d\vec{B}$ are given by the above eqns. (55) and (56), respectively, but with $q$ replaced by $dq$. The net force $F$ on the upper line charge due to lower line charge is given by

$$F = \int_{-L/2}^{L/2}\int_{-L/2}^{L/2} dq'\left(d\vec{E} + \vec{v}\times d\vec{B}\right) =$$

$$= \int_{-L/2}^{L/2}\int_{-L/2}^{L/2} dq'\left(d\vec{E} + \frac{1}{c^2}\,\vec{v}\,\times\left(\vec{v}\,\times d\vec{E}\right)\right) =$$

$$= \int_{-L/2}^{L/2}\int_{-L/2}^{L/2} dq'\left(1 - \frac{v^2}{c^2}\right)d\vec{E} + \frac{1}{c^2}\int_{-L/2}^{L/2}\int_{-L/2}^{L/2} dq'\,\vec{v}\left(\vec{v}\cdot d\vec{E}\right) . \tag{57}$$

In the last step we have used a well-know identity for the triple vector product. We now let the vertical direction be the $x$ direction, and we resolve $d\vec{E}$ into its $z$ and $x$ components:

$$d\vec{E} = dE_x\,\hat{i} + dE_z\,\hat{k} ,$$

$$dE_x = dE\sin\theta \quad , \quad dE_z = dE\cos\theta \quad ,$$

$$dE = \frac{1}{4\pi\epsilon_0}\frac{dq\left(1 - \frac{v^2}{c^2}\right)}{r_0^2\left(1 - \frac{v^2}{c^2}\sin^2\theta\right)^{3/2}} . \tag{58}$$

The angle $\theta$ is now the angle between the positive $z$ axis and the line connecting the infinitesmal elements of charge $dq$ and $dq'$. We may resolve $\vec{F}$ into its $x$ and $z$ components i.e. $\vec{F} = F_x \hat{i} + F_z \hat{k}$. For $F_x$ we have

$$F_x = \int_{-L/2}^{L/2} \int_{-L/2}^{L/2} dq' \left(1 - \frac{v^2}{c^2}\right) dE_y =$$

$$= \int_{-L/2}^{L/2} \int_{-L/2}^{L/2} dq' \left(1 - \frac{v^2}{c^2}\right) dE \sin\theta =$$

$$= \frac{1}{4\pi\epsilon_0} \int_{-L/2}^{L/2} \int_{-L/2}^{L/2} dq' \left(1 - \frac{v^2}{c^2}\right) \frac{dq \left(1 - \frac{v^2}{c^2}\right)}{r_0^2 \left(1 - \frac{v^2}{c^2}\sin^2\theta\right)^{3/2}} \sin\theta =$$

$$= \frac{1}{4\pi\epsilon_0} \left(1 - \frac{v^2}{c^2}\right)^2 \int_{-L/2}^{L/2} \int_{-L/2}^{L/2} dq' \frac{dq}{r_0^2 \left(1 - \frac{v^2}{c^2}\sin^2\theta\right)^{3/2}} \sin\theta . \quad (59)$$

Using $dq = \lambda \, dz$, $dq' = \lambda \, dz'$, $r_0 = \sqrt{(z-z')^2 + r^2}$, and $\sin\theta = \frac{r}{r_0}$ in this equation yields

$$F_x = \frac{\lambda^2}{4\pi\epsilon_0} \left(1 - \frac{v^2}{c^2}\right)^2 \int_{z'=-L/2}^{L/2} \int_{z=-L/2}^{L/2} dz' \, dz \, \frac{r_0^3}{r_0^2 \left(r_0^2 - \frac{v^2}{c^2}r^2\right)^{3/2}} \frac{r}{r_0} =$$

$$= \frac{\lambda^2}{4\pi\epsilon_0} \left(1 - \frac{v^2}{c^2}\right)^2 r \int_{z'=-L/2}^{L/2} \int_{z=-L/2}^{L/2} dz' \, dz \, \frac{1}{\left(r_0^2 - \frac{v^2}{c^2}r^2\right)^{3/2}} =$$

$$= \frac{\lambda^2}{4\pi\epsilon_0} \left(1 - \frac{v^2}{c^2}\right)^2 r \int_{z'=-L/2}^{L/2} \int_{z=-L/2}^{L/2} dz' \, dz \, \frac{1}{\left((z - z')^2 + \left(1 - \frac{v^2}{c^2}\right)r^2\right)^{3/2}} =$$

$$= \frac{\lambda^2 L}{2\pi\epsilon_0 r} \left(1 - \frac{v^2}{c^2}\right)^{3/2} \left(\sqrt{\frac{1}{1 - \frac{v^2}{c^2}} + \frac{r^2}{L^2}} - \frac{r}{L}\right) . \quad (60)$$

It is clear from the last term of this equation that $F_x$ agrees with eqn. (9) up to terms of lowest order in $r/L$, and, we leave it to the reader to convince himself that, just as in the static case, $F_z$ vanishes by symmetry.

In concluding this section we point out that a problem very similar ours is treated in Feynman's famous monograph, the *Feynman Lectures on Physics* [20]. Feynman considers, from a relativistic standpoint, the interaction of an infinitely long, straight current carrying wire with a negatively charged particle moving with uniform velocity in a direction parallel to the wire. The velocity of the particle is the same as the velocity of the conduction electrons in the wire. Just as is done here, he analyses the situation in two different reference frames, one fixed with respect to the the wire, so that the wire is at rest in that frame, and another reference frame in which the particle is at rest. In the first reference frame there is only a magnetic force on the particle, but in the reference frame in which the particle is at rest it seems at first glance that there is no force on the particle. Feynman argues that the only way out of this paradox is to agree that "a neutral wire with a current appears to be charged when set in motion." [20] He then goes on to describe the difference in the motions of the positive and negative charge carriers in the current carrying wire, and from this analysis together with length contraction he is able to calculate the charge density of the moving wire, and, consequently the electric force on the electron. It turns out, of course, to be exactly the correct one necessary to resolve the paradox.

## 5. Conclusions

Everything which was used above in order to get at length contraction, even the Lorentz force law*, was well-known to the Maxwellians years before 1887, the year of the Michelson-Morley experiment, and the arguments presented above give additional support to Redžić's contention that the Maxwellians could perhaps have arrived at the notion of length contraction much earlier than 1887. In fact, there are even more trivial examples which come about from trying to uphold the principle of relativity in electromagnetism. Consider the example presented in the opening paragraph of Einstein's 1905 paper, namely, that of a charged particle near a magnet. If the charge is stationary relative to the magnet's rest frame, then according to the Lorentz force law, there should be no magnetic force on the charge; however, if it is moving relative to the magnet in a direction which is different than the direction of the magnetic field, so that $\vec{v} \times \vec{B}$ is not zero, then there is a non-zero force on the charge. This simple experiment with a charge and a magnetic enables us to discern whether or not we are moving in uniform translation relative to a fixed observer, again in violation of the principle of relativity. In our view, it seems almost impossible to believe that such difficulties as the ones described in this paper and in [1] could have gone unnoticed by the Maxwellians, and our results give even more support to Redžić's contention that the Maxwellians could have come up with the contraction hypothesis much earlier. In fact, according to Everitt, "asymmetries that do not appear to be inherent in the phenomena"♯ such as the ones described here did not escape the genius of Maxwell, who, "had he not died so young . . . would almost certainly have developed special relativity a decade or more before Einstein" [21]. Everitt is almost certain that he can trace the first use of "relativity" in the modern sense as follows: [22], [24]

$$\text{Maxwell} \longrightarrow \text{Poincaré} \longrightarrow \text{Einstein} .$$

---

* According to Whittaker ([19]), the first person to obtain the Lorentz force law, essentially in its present form, was J.J. Thompson in 1881, two years after Maxwell's death, except that he had an error of 1/2 in the magnetic term. Thompson assumed that the velocity of the charge in the Lorentz force law was not relative to the magnetic field but relative to the medium (or "aether"). Eight years later Heaviside corrected the factor of 1/2. In 1895, Lorentz first wrote down this law which now bears his name. However, it seems not to be mentioned in Whittaker that Maxwell had a version of it which appeared in his *Treatise on Electricity and Magnetism* (1865)[21].

♯ Specifically, in the original German i.e. in the first sentence of Einstein's 1905 *Annalen der Physik* article [23]), it reads: "... Asymmetrien ..., welche den Phänomenen nicht anzuhaften scheinen ...."

One finds, according to Everitt, the use of the word *relativity*, in the present-day sense, in Sections 105 and 106 of Maxwell's book, *Matter and Motion*, published in 1877 [25]. To lend support to Everitt's conclusion, we point out that Poincaré makes acknowledgement in his book *Science and Hypothesis* to having devoted considerable time in mastering Maxwell's ideas [26], and further support for this can be found in [27], [28], [29].

Finally, we point out that in the above analysis of two parallel line charges not only have we explained an anomaly in classical electromagnetism, but also ours is much simpler than Redzic's example, while still retaining, in so far as it relates to length contraction, all of the essential features of his more complicated example. Additionally, it possesses sufficient symmetry so as to permit us to go a long way towards actually getting at the Lorentz contraction formula. In contrast, Redzic only points out the existence of an anomaly about electromagnetic images of moving and stationary charge distributions. One could, in fact, imagine trying to carry out in Redzic's case a development paralleling ours. After all, electric field lines and equipotential surfaces along with electromagnetic images must, if logically analysed to their empirical origins, ultimately involve forces and charges in their descriptions.

## Acknowledgements

Useful discussions with Dr. Mikhail Kagan are gratefully acknowledged.

## References

[1] Redžić, D V 2004 Image of a moving sphere and the FitzGerald-Lorentz contraction *Eur. J. Phys.* **25** 123-6
[2] Poincaré H 1904 "L'état actuel et l'avenir de la physique mathématique" *Bulletin des Sciences mathématiques*, 28, 2nd series (reorganized 39-1), pages 302-324 (the English translation by George Bruce Halsted was published in *The Monist* on January, 1905).
[3] Redžić, D V 2002a *Am. J. Phys* **60** 275-7
[4] Redžić, D V 2002b *Am. J. Phys* **60** 506-8
[5] Hunt, B J 1991, The Maxwellians, Cornell University Press, Ithaca, New York, USA
[6] Michelson A A 1881 The relative motion of the Earth and the lumineferous ether *Am. J. Sci* **22** 20
[7] Michelson A A, Morley E W (1887), "On the Relative Motion of the Earth and the Luminiferous Ether", American Journal of Science 34, 333-345.
[8] FitzGerald G F 1889 The ether and the Earth's atmosphere *Science* **13** 390
[9] Lorentz H A 1895, Versuch einer Theorie der electrischen und optischen Erscheinungen in bewegten Körpern, Brill, Leyden.
[10] Brush S G.1967 Note on the History of the FitzGerald-Lorentz Contraction *Isis* **58** , 230-232
[11] George E. A. Matsas 2003 *Phys. Rev. D* **68** 027701.
[12] Moeller H C 1972 The Theory of Relativity, 2nd Edition, Oxford, Clarendon Press

[13] Poincaré H. 1905 Sur la dynamique de l'electron *Compte-rendus de l'Acad. Scien.*, **140**, p. 1505, eqn. (2). (The composition of velocities formula (eqn. (40)) first appeared in this note of Poincaré and then subsequently in the 1905 article of Einstein referenced below and again by Poincaré in: Poincaré, H. 1906 Sur la dynamique de l'electron *Rendiconti del Circolo Matematico di Palermo*, **21**, 129-176.)

[14] Martins R 1982 *Am J. Phys.* **50** 1008-1011.

[15] Lorrain P, Corson D R 1970 Electromagnetic Fields and Waves, 2nd Edition, W.H. Freeman, New York, p. 272 to 275 .

[16] Tipler Paul A, Mosca Gene P 2007 Physics for Scientists and Engineers, 6th Edition, W.H. Freeman, New York.

[17] Belliostin S B 1971 Classical Electron Theory, Vishnaiya Skola, Moskva, pp. 156-161 (in Russian).

[18] Jackson J D 1972 Classical Electrodynamics, 2nd Edition, Wiley,, New York [compare p. 555, eqn. (11.154) for $\vec{E}$ (our eqn. (55)) and p. 542, eqn. (11.102) for $\vec{B}$ (our eqn. (56))].

[19] Whittaker E T 1987 A History of the Theories of Aether and Electricity, American Institute of Physics, New York , vol. 1.

[20] Feynman R P, Leighton R B, Sands M 1964 The Feynman Lectures on Physics, Addison-Wesley, Reading, MA, USA, Vol. 2, pp. 13-6 to 13-11.

[21] Everitt C W Francis 2006 "James Clerk Maxwell: a force for physics" *Physics World*: http://physicsworld.com/cws/article/print/26527 (online publication).

[22] Everitt C W Francis 2008 "Kelvin, Maxwell, Einstein and the Ether - Who was Right about What?" in *Kelvin Life, Labours and Legacy*, edited by Raymond Flood, Mark McCartney and Andrew Whitaker, Oxford University Press, Oxford, England.

[23] Einstein A 1905 "Zur Elektrodynamik bewegter Körper", *Ann. der Phys.*, 17, 891-921.

[24] personal communication with Francis Everitt.

[25] Maxwell James Clerk 1952 *Matter and Motion*, Dover, New York.

[26] Poincaré H 1902 *La Science et l'hypothese*, Flammarion, Paris (English translation by John W. Bolduc: Science and Hypothesis, Dover, New York, 2003).

[27] Keswani G H 1965 "Origin and Concept of Relativity (I)", *The British Journal for the Philosophy of Science*, 15, 60 pp. 286-306.

[28] Darrigol O 2004 "The Mystery of the Einstein-Poincaré Connection" *Isis* 95, 614-626.

[29] Coleman B (2003) *Eur. J. Phys.* **24** 301-313; (2004)*Eur. J. Phys.* **25**L31